\documentclass[preprintnumbers,10pt,nofootinbib]{revtex4}
\pdfoutput=1

\usepackage{amsmath,latexsym,amssymb,amsfonts}
\usepackage[pdftex]{color,graphicx}
\usepackage{bm}

\begin{document}

\title{\textbf{Gravitational waves from cosmic bubble collisions}}
\author{\textsc{Dong-Hoon Kim}$^{a,b}$\footnote{{\tt ki13130@gmail.com}}, \textsc{Bum-Hoon Lee}$^{d,e}$\footnote{{\tt bhl@sogang.ac.kr}}, \textsc{Wonwoo Lee}$^{d}$\footnote{{\tt warrior@sogang.ac.kr}}, \textsc{Jongmann Yang}$^{a,b,c}$\footnote{{\tt jyang@ewha.ac.kr}}\\
and \textsc{Dong-han Yeom}$^{d,f,g}$\footnote{{\tt innocent.yeom@gmail.com}}}
\affiliation{\textit{$^{a}$\small{Basic Science Research Institute, Ewha Womans
University, Seoul 120-750, Republic of Korea}}\\
\textit{$^{b}$\small{Institute for the Early Universe, Ewha Womans
University, Seoul 120-750, Republic of Korea}}\\
\textit{$^{c}$\small{Department of Physics, Ewha Womans University, Seoul
120-750, Republic of Korea}}\\
\textit{$^{d}$\small{Center for Quantum Spacetime, Sogang University,
Seoul 121-742, Republic of Korea}}\\
\textit{$^{e}$\small{Department of Physics, Sogang University, Seoul
121-742, Republic of Korea}}\\
\textit{$^{f}$\small{Yukawa Institute for Theoretical Physics, Kyoto University, Kyoto
606-8502, Japan}}\\
\textit{$^{g}$\small{Leung Center for Cosmology and Particle Astrophysics,
National Taiwan University, Taipei 10617, Taiwan}}}

\begin{abstract}
Cosmic bubbles are nucleated through the quantum tunneling process. After
nucleation they would expand and undergo collisions with each other. In this
paper, we focus in particular on collisions of two equal-sized bubbles and
compute gravitational waves emitted from the collisions. First, we study the
mechanism of the collisions by means of a real scalar field and its quartic
potential. Then, using this model, we compute gravitational waves from the
collisions in a straightforward manner. In the quadrupole approximation, 
\textit{time-domain} gravitational waveforms are directly obtained by
integrating the energy-momentum tensors over the volume of the wave sources,
where the energy-momentum tensors are expressed in terms of the scalar
field, the local geometry and the potential. We present gravitational
waveforms emitted during (i) the initial-to-intermediate stage of strong
collisions and (ii) the final stage of weak collisions: the former is
obtained numerically, in \textit{full General Relativity} and the latter
analytically, in the flat spacetime approximation. We gain \textit{%
qualitative} insights into the time-domain gravitational waveforms from
bubble collisions: during (i), the waveforms show the non-linearity of the
collisions, characterized by a modulating frequency and cusp-like bumps,
whereas during (ii), the waveforms exhibit the linearity of the collisions,
featured by smooth monochromatic oscillations.
\end{abstract}

\maketitle

\newpage

\tableofcontents



\section{Introduction\label{intro}}

A detection of signatures of primordial gravitational waves (GWs) in the
cosmic microwave background was claimed by the BICEP2 experiment in 2014~%
\cite{BICEP2}. But this was shown to be likely caused by interstellar dust
soon thereafter \cite{Flauger}, and the search for true signatures of
primordial GWs is still ongoing. The detection of the GW signatures, if
confirmed, would gain the greatest importance, among other things, from its
link to cosmic \textquotedblleft inflation\textquotedblright : primordial
GWs are seen as the smoking gun for the \textquotedblleft Big
Bang\textquotedblright\ expansion. According to the inflation theory, the
early universe experienced an extreme burst of expansion, which lasted a
tiny fraction of a second, but smoothed out irregularities--inhomogeneities,
anisotropies and the curvature of space, and made the universe appear
homogeneous and isotropic~\cite{Guth-Linde-Albrecht}.

It has been suggested that inflationary models of the early universe most
likely lead to a \textquotedblleft multiverse\textquotedblright ~\cite%
{Linde-Guth}. One such model is \textquotedblleft eternal
inflation\textquotedblright ~\cite{vilen}: it proposes that many bubbles of
spacetime individually nucleate and grow inside an ever-expanding background
multiverse. The nucleation and growth of such bubbles can be modeled by a
Coleman-de Luccia (CDL) instanton, a type of quantum transition between two
classically disconnected vacua at different energies; the higher energy
(false vacuum), the lower energy (true vacuum)~\cite{CDL-BHLWL}. A scalar
field initially in the false vacuum state may tunnel quantum mechanically to
the true vacuum state. This nucleates bubbles of the true vacuum (new phase)
inside of the false vacuum (old phase) background; through a first-order
phase transition. These bubbles then expand and collide with each other. The
mechanism of bubble collisions can be effectively modeled by the CDL
instanton: as bubbles continue to collide repeatedly, the scalar field
transitions back and forth repeatedly between the false vacuum and the true
vacuum, eventually settling down in the true vacuum as the collision process
is gradually terminated.

From the viewpoints of physical cosmology, bubble collisions and GWs emitted
from the collisions are interesting in the following contexts: (1) Our
primordial inflation would be completed by a second-order (not by a
first-order) phase transition. However, there is a possibility that some
weaker inflation could occur after the primordial inflation; for example,
\textquotedblleft thermal inflation\textquotedblright\ \cite{Lyth}. It is
quite probable that the thermal inflation is completed by a first-order
phase transition, and therefore bubble collisions could take place through a
CDL instanton. Then there would be some signatures of bubble collisions,
which would presumably be carried via GWs \cite{Easther}. (2) Suppose that
we live in a single large true vacuum bubble and that the boundary of our
bubble would collide with another bubble that is located outside our
observable universe \cite{Chang}. Then there would exist some signatures of
bubble collisions and these could be carried via GWs. For scenario (1), the
mechanism of bubble collision - GW emission should be modeled
stochastically. However, for scenario (2), the mechanism can be well
approximated by a two-bubble collision model.

There were numerous studies about bubble collisions and GWs emitted from the
collisions. Among others, Hawking \textit{et al}.~\cite{Hawking} and Wu~\cite%
{Wu} studied the mechanism of the collision of two bubbles using the
thin-wall approximation. Johnson \textit{et al}.~\cite{Johnson} and Hwang 
\textit{et al}.~\cite{Hwang} investigated the collision of two bubbles in
full General Relativity via numerical computations. Kosowsky \textit{et al}. 
\cite{Kosowsky} computed the GW spectrum resulting from two-bubble
collisions in first-order phase transitions in flat spacetime using
numerical simulations. Caprini \textit{et al}.~\cite{Caprini} developed a
model for the bubble velocity power spectrum to calculate analytically the
GW spectrum generated by two-bubble collisions in first-order phase
transitions in flat spacetime.

In this paper, we focus on collisions of two equal-sized bubbles and compute
GWs emitted from the collisions in \textit{time domain}. Largely, our
analysis proceeds in two steps through Sections~\ref{geometry} and \ref{GW}.
In Section~\ref{geometry}, we study the mechanism of bubble collisions by
means of a real scalar field and a quartic potential of this field, building
the simplest possible model for a CDL instanton. Einstein equations and a
scalar field equation are derived for this system and are solved
simultaneously for the full General Relativistic treatment of the collision
dynamics. Hwang \textit{et al}.~\cite{Hwang} is closely reviewed for this
purpose. In Section~\ref{GW}, using the scalar field model from Section~\ref%
{geometry}, we compute GWs from the bubble collisions in a straightforward
manner. In the quadrupole approximation, time-domain gravitational waveforms
are directly obtained by integrating the energy-momentum tensors over the
volume of the wave sources, where the energy-momentum tensors are expressed
in terms of the scalar field, the local geometry and the potential;
therefore, containing all necessary information about the bubble collisions.
Part of computational results from Ref.~\cite{Hwang} is recycled here to
build the energy-momentum tensors. In parallel with the scalar field
solutions in Section~\ref{geometry}, which have been obtained with various
false vacuum field values~\cite{Hwang}, we present gravitational waveforms
emitted during (i) the initial-to-intermediate stage of strong collisions
and (ii) the final stage of weak collisions: the former is obtained
numerically, in \textit{full General Relativity} and the latter
analytically, in the flat spacetime approximation. The thin-wall and
quadrupole approximations are assumed to simplify our analysis and the
next-to-leading order corrections beyond these approximations are
disregarded. However, the approximations serve our purpose well: we aim to
provide \textit{qualitative} illustrations of the time-domain gravitational
waveforms from the bubble collisions, which will be useful for constructing
the templates for observation in the future. We adopt the unit convention, $%
c=G=1$ for all our computations of GWs.

\section{Gravity-scalar field dynamics for colliding bubbles\label{geometry}}

The mechanism of two equal-sized colliding bubbles can be effectively
modeled by means of a CDL instanton~\cite{CDL-BHLWL}. Basically, one can
build a model for this, which consists of gravitation, a real scalar field
and a potential of the field. In this Section we introduce one such model
from Hwang \textit{et al}.~\cite{Hwang}, which is built with a quartic
potential, the simplest possible one for the CDL instanton.

\subsection{Dynamics of bubble collisions \label{basics}}

A system of Einstein gravity coupled with a scalar field that is governed by
a potential can be described by the following action: 
\begin{equation}
\mathcal{S}=\int d^{4}x\sqrt{-g}\left[\frac{1}{16\pi}R-\frac{1}{2}%
\nabla_{\mu}\phi\nabla^{\mu}\phi-V\left(\phi\right)\right] \,,
\end{equation}
where $R$ denotes the Ricci scalar, $\phi$ the scalar field and $%
V\left(\phi\right)$ the potential of the scalar field~\cite{Hwang}. From
this system the Einstein equations are derived: 
\begin{equation}
R_{\mu\nu}-\frac{1}{2}Rg_{\mu\nu}=8\pi T_{\mu\nu},  \label{ein}
\end{equation}
where the energy-momentum tensors on the right-hand side are written as 
\begin{equation}
T_{\mu\nu}=\phi_{;\mu}\phi_{;\nu}-\frac{1}{2}\phi_{;\rho}\phi_{;\sigma}g^{%
\rho\sigma}g_{\mu\nu} -V\left(\phi\right)g_{\mu\nu}\,.  \label{T}
\end{equation}
Also, the scalar field equation for the system reads 
\begin{equation}
\nabla^{2}\phi=\frac{dV}{d\phi}\,.  \label{scl}
\end{equation}

The Einstein equations~(\ref{ein}) and the scalar field equation~(\ref{scl})
constitute a scalar field model that effectively describes the mechanism of
two colliding bubbles in curved spacetime~\cite{Hwang}. Given a potential $%
V\left( \phi \right) $, the scalar field solution $\phi $ and the geometry
solution $g_{\mu \nu }$ should be obtained by solving Eqs.~(\ref{ein}) and (%
\ref{scl}) simultaneously \footnote{%
This inevitably results in the effects of radiation reaction being included
in the solutions, $\phi $ and $g_{\mu \nu }$~\cite{Hwang}.}. To this end, we
prescribe an ansatz for the geometry $g_{\mu \nu }$ with the hyperbolic
symmetry, using the \textit{double-null} coordinates: 
\begin{equation}
ds^{2}=-\alpha _{\mathrm{h}}^{2}\left( u,v\right) dudv+r_{\mathrm{h}%
}^{2}\left( u,v\right) dH^{2}\,,  \label{dsgen}
\end{equation}%
where $dH^{2}=d\chi ^{2}+\sinh ^{2}\chi d\theta ^{2}$ with $0\leq \chi
<\infty $, $0\leq \theta <2\pi$~\cite{Hwang}, and $\alpha _{\mathrm{h}%
}\left( u,v\right) $ and $r_{\mathrm{h}}\left( u,v\right) $ are to be
determined by solving Eqs.~(\ref{ein}) and (\ref{scl}) simultaneously in the
coordinates $\left( u,v,\chi ,\theta \right) $. In the flat spacetime limit,
the double-null coordinates are defined as $u\equiv \tau -x$ and $v\equiv
\tau +x$ with $\tau ^{2}\equiv t^{2}-y^{2}-z^{2}$, $t=\tau \cosh \chi $, $%
y=\tau \sinh \chi \sin \theta $, $z=\tau \sinh \chi \cos \theta $: in our
analysis, the $x$-axis of Cartesian coordinates is chosen to coincide with a
line adjoining the centers of the two bubbles, and the $y$-axis and the $z$%
-axis lie in a plane perpendicular to the $x$-axis.

\subsection{Solving the scalar field equation\label{scalar}}

To build the simplest model of the CDL instanton for two identical colliding
bubbles, we consider the potential in Subsection~\ref{basics} to be 
\begin{equation}
V\left( \phi \right) =\frac{p_{4}}{4!}\phi ^{4}+\frac{p_{3}}{3!}\phi ^{3}+%
\frac{p_{2}}{2!}\phi ^{2}\,,  \label{V1}
\end{equation}%
where $p_{2}$, $p_{3}$ and $p_{4}$ are constants which can be appropriately
chosen to tune the shape of the potential. Bubble collisions are represented
by the scalar field moving along this potential: the field initially in the
false vacuum state (at higher local minimum of potential) tunnels quantum
mechanically to the true vacuum state (at lower local minimum of potential),
repeating the transitions back and forth between the two states, eventually
settling down in the true vacuum state.

Following Ref.~\cite{Hwang}, we may rescale the scalar field, $S\equiv \sqrt{%
4\pi }\phi $ for computational convenience, and can specify $p_{2}$, $p_{3}$
and $p_{4}$ in terms of the false vacuum field $S_{\mathrm{f}}=\sqrt{4\pi }%
\phi _{\mathrm{f}}$, the vacuum energy of the false vacuum $V_{\mathrm{f}}$
and a free parameter $\beta $. The potential in Eq.~(\ref{V1}) can then be
rewritten as 
\begin{equation}
V\left( S\right) =\frac{3V_{\mathrm{f}}}{\beta S_{\mathrm{f}}^{4}}S^{4}-%
\frac{2\left( \beta +3\right) V_{\mathrm{f}}}{\beta S_{\mathrm{f}}^{3}}S^{3}+%
\frac{3\left( \beta +1\right) V_{\mathrm{f}}}{\beta S_{\mathrm{f}}^{2}}%
S^{2}\,.  \label{V}
\end{equation}%
With this potential the scalar field equation~(\ref{scl}), which is now
rescaled, reads 
\begin{equation}
\nabla ^{2}S=\frac{12V_{\mathrm{f}}}{\beta S_{\mathrm{f}}^{4}}S^{3}-\frac{%
6\left( \beta +3\right) V_{\mathrm{f}}}{\beta S_{\mathrm{f}}^{3}}S^{2}+\frac{%
6\left( \beta +1\right) V_{\mathrm{f}}}{\beta S_{\mathrm{f}}^{2}}S\,.
\label{scl1}
\end{equation}%
This is a non-linear wave equation whose analytical solution is not
generally known: we normally approach this type of problem with numerical
methods.

\begin{figure}[tbp]
\begin{center}
\includegraphics[scale=0.75]{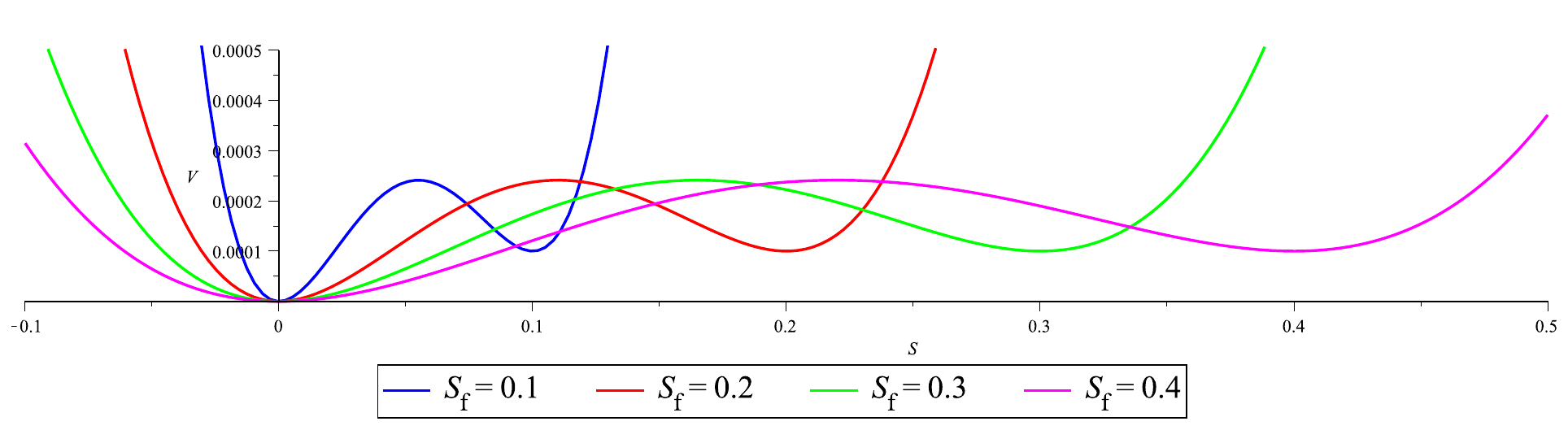}
\end{center}
\caption{(color online). The quartic potential $V\left( S\right) =\frac{3V_{%
\mathrm{f}}}{\protect\beta S_{\mathrm{f}}^{4}}S^{4}-\frac{2\left( \protect%
\beta +3\right) V_{\mathrm{f}}}{\protect\beta S_{\mathrm{f}}^{3}}S^{3}+\frac{%
3\left( \protect\beta +1\right) V_{\mathrm{f}}}{\protect\beta S_{\mathrm{f}%
}^{2}}S^{2}$; expressed in terms of the false vacuum field $S_{\mathrm{f}}=%
\protect\sqrt{4\protect\pi }\protect\phi _{\mathrm{f}}$, the vacuum energy
of the false vacuum $V_{\mathrm{f}}$ and a free parameter $\protect\beta $.
The patterns of the potential are shown for $\protect\beta =0.1$, $V_{%
\mathrm{f}}=10^{-4}$ and $S_{\mathrm{f}}=\protect\sqrt{4\protect\pi }\protect%
\phi _{\mathrm{f}}=$ (1) $0.1$, (2) $0.2$, (3) $0.3$, (4) $0.4$. (Credit:
Hwang \textit{et al.}~\protect\cite{Hwang})}
\label{fig1}
\end{figure}

\begin{figure}[tbp]
\begin{center}
\includegraphics[scale=0.2]{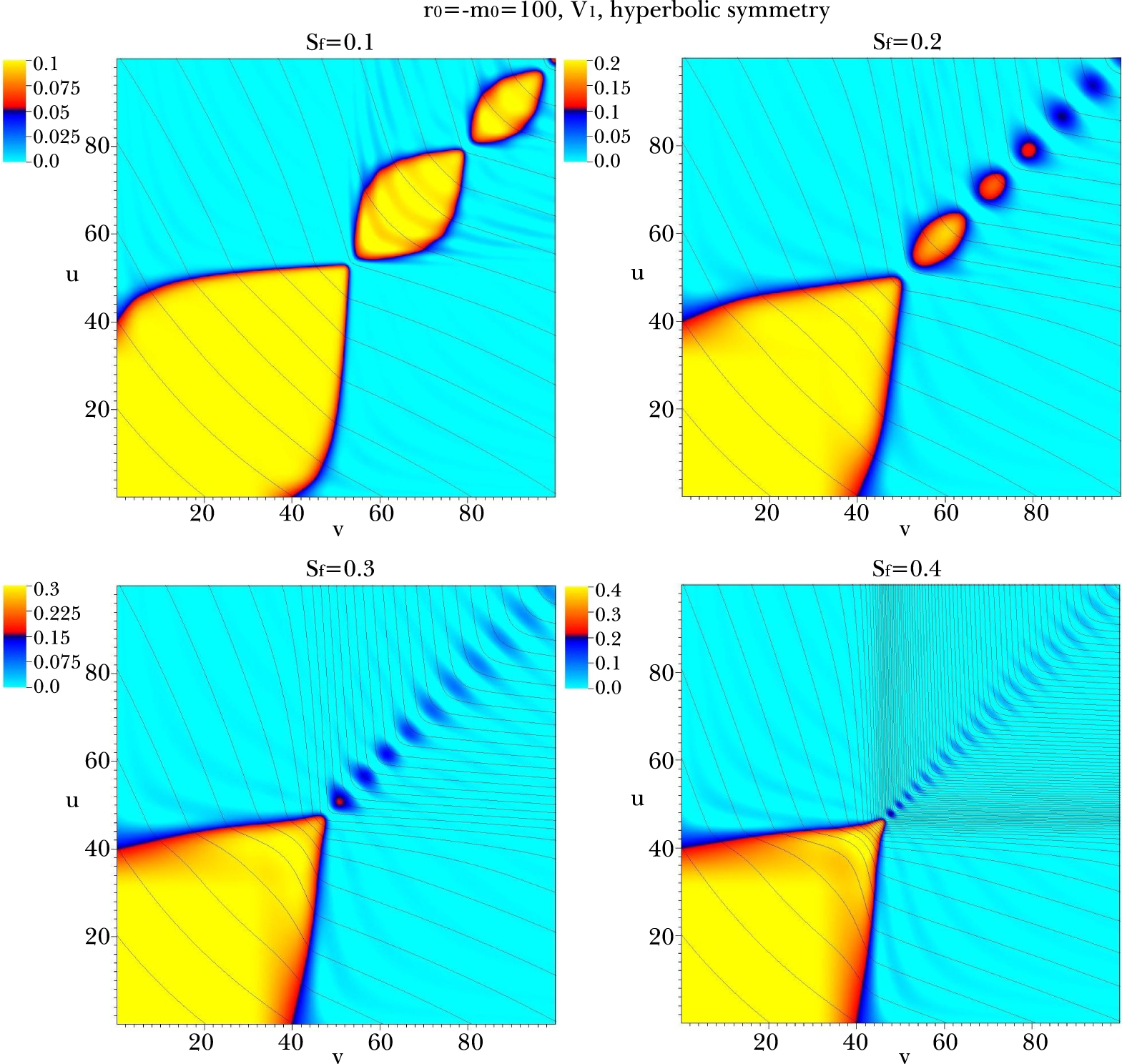}
\end{center}
\caption{(color online). The numerical solutions $S\left(u,v\right)=\protect%
\sqrt{4\protect\pi}\protect\phi\left(u,v\right)$ obtained with the potential
from Figure~\protect\ref{fig1}; with various false vacuum field values, $S_{%
\mathrm{f}}=\protect\sqrt{4\protect\pi}\protect\phi_{\mathrm{f}}=$ (1) $0.1$%
, (2) $0.2$, (3) $0.3$, (4) $0.4$. (Credit: Hwang \textit{et al.}~%
\protect\cite{Hwang})}
\label{fig2}
\end{figure}

Now, we solve the scalar field equation~(\ref{scl1}) simultaneously with the
Einstein equations~(\ref{ein}), using the ansatz given by~Eq. (\ref{dsgen}),
in the coordinates $\left( u,v,\chi ,\theta \right) $. However, it turns out
that our scalar field solution is independent of the coordinates $\chi $ and 
$\theta $ and is expressed in the coordinates $\left( u,v\right) $ only;
namely, $S\left( u,v\right) =\sqrt{4\pi }\phi \left( u,v\right) $~\cite%
{Hawking, Hwang, Kosowsky}. With the choice of the constants, $\beta =0.1$, $%
V_{\mathrm{f}}=10^{-4}$ and $S_{\mathrm{f}}=$ (1) $0.1$, (2) $0.2$, (3) $0.3$%
, (4) $0.4$ in Eq.~(\ref{V}), the potential takes the forms as given by
Figure~\ref{fig1}~\cite{Hwang}. With this potential, our numerical solution $%
S\left( u,v\right) $ is obtained as presented by Figure~\ref{fig2}~\cite%
{Hwang}.{\ In each case of }$S_{\mathrm{f}}${, (1) - (4), the bubble wall
has a different value of tension due to a different value of }$S_{\mathrm{f}%
} ${\ as shown by Figure~\ref{fig1}. In the top left of Figure \ref{fig2}
the bubble has the lowest tension while in the bottom right it has the
highest tension among the four cases of }$S_{\mathrm{f}}${. This results in
the wall crossing regions in the top left being relatively wider than those
in the bottom right. }

\section{Gravitational waves from bubble collisions\label{GW}}

\begin{figure}[tbp]
\begin{center}
\includegraphics[scale=0.75]{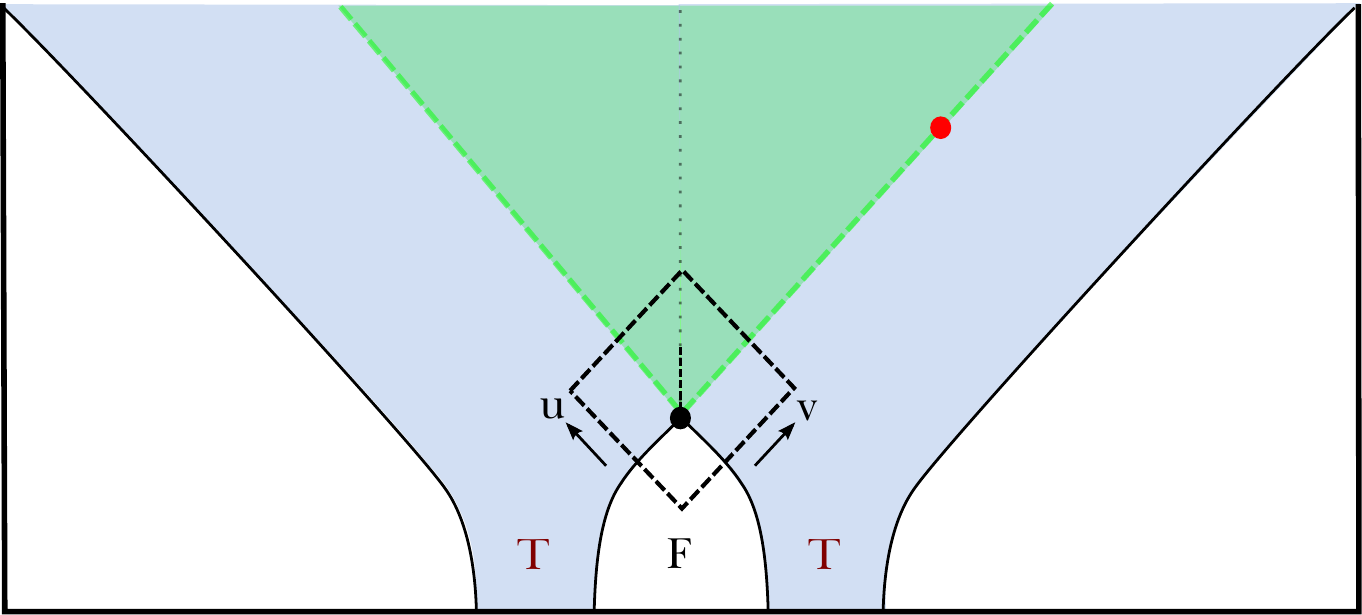}
\end{center}
\caption{(color online). A causal relationship between two colliding bubbles
and an observer: Gravitational waves from bubble collisions can be observed
by a distant observer (red spot) who is in the causal future of a collision
event (black spot). The dotted vertical line inside the timelike zone
(green-colored region) of a null-cone (green dashed lines) represents the
time-axis, along which bubble collisions took place in the distant past;
within the intersection of the timelike zone of the null-cone and the
`diamond' zone (region enclosed by black dashed lines). $u$ and $v$
represent the double-null coordinates as defined in Section \protect\ref%
{geometry}. T and F denote a true vacuum and a false vacuum, respectively. }
\label{fig3}
\end{figure}

In Section \ref{geometry} we have built a system of two equal-sized
colliding bubbles in curved spacetime by means of a CDL instanton model,
considering the potential given by Eq. (\ref{V}) \cite{Hwang}. Now, we
consider that the a present observer lives in the true vacuum region of one
of the two bubbles and that signatures of bubble collisions which took place
in the distant past are being carried to the present observer via GWs. Here
we assume that the distance between the center of collision region and the
observer can be arbitrarily large (within the size of our universe), and
thus that the time for a collision event, which is the retarded time to the
present observer, can be quite far in the distant past; namely, $t_{\mathrm{R%
}}=t_{\mathrm{P}}-r/c\ll t_{\mathrm{P}}$, where $t_{\mathrm{R}}$ denotes the
retarded time, $t_{\mathrm{P}}$ the present time and $r$ the distance.
Therefore, our bubble collision may be regarded as a localized event, as
long as the observer is reasonably far away from the collision region. In
Figure \ref{fig3} a causal relationship between two colliding bubbles and an
observer is depicted using a null-cone. Here collision events that took
place in the distant past are placed within the intersection of the timelike
zone (green-colored region) of a null-cone (green dashed lines) and the
`diamond' zone (region enclosed by black dashed lines): all the collision
events as our GW sources, namely the collisions in {Figure \ref{fig2}
(strong collisions in the initial-to-intermediate stage) and the collisions
to be discussed in Subsection \ref{alter} later (weak collisions in the
final stage) should be considered to have taken place within this
intersection \footnote{%
In principle, the diamond zone can be extended to cover the longer evolution
of bubble collision. This will result in the larger intersection area with
the timelike zone of the null-cone. However, no matter how large the
intersection area is, it should still be regarded as a well-localized region
for our GW sources: we assume that a present observer is reasonably far away
from the sources in our computations of GWs. This naturally renders our
results convergent. While we consider only the retarded field for our
`time-domain' GWs, Kosowsky\textit{\ et al. }\cite{Kosowsky} take both the
retarded and advanced fields for their `frequency-domain' GWs. On account of
this, their computation domain is unbounded, but they obtain convergent
results using a method of `phenomenological cutoff'.}.}

In the above scenario, our GWs from bubble collisions can be computed in a
straightforward manner: by integrating the energy-momentum tensors combined
with a Green's function over the volume of the wave sources, where the
energy-momentum tensors are expressed in terms of the scalar field, the
local geometry and the potential by means of Eqs. (\ref{T}), (\ref{dsgen})
and (\ref{scl1}); therefore, containing all necessary information about the
bubble collisions. A mathematical description of this computation is given
as follows. In the transverse trace-free (TT) gauge, GWs, as derived from
the perturbed Einstein equations in linearized gravity, can be expressed as 
\begin{equation}
h_{ij}^{\mathrm{TT}}\left( t,\mathbf{x}\right) =\overline{h}_{ij}^{\mathrm{TT%
}}\left( t,\mathbf{x}\right) =\frac{4G}{c^{4}}\Lambda _{ij,kl}\left( \mathbf{%
n}\right) \int d^{3}x^{\prime }\frac{1}{\left\vert \mathbf{x-x}^{\prime
}\right\vert }T^{kl}\left( t-\frac{\left\vert \mathbf{x-x}^{\prime
}\right\vert }{c},\mathbf{x}^{\prime }\right) \,,  \label{h0}
\end{equation}%
where $\overline{h}_{ij}\equiv h_{ij}-\frac{1}{2}\delta _{ij}h_{k}^{k}$ and
the unit vector $\mathbf{n}$ denotes the propagation direction of the waves,
and the projection tensor for gravitational radiation, 
\begin{equation}
\Lambda _{ij,kl}\left( \mathbf{n}\right) \equiv P_{ik}P_{jl}-\frac{1}{2}%
P_{ij}P_{kl}  \label{L}
\end{equation}%
with 
\begin{equation}
P_{ij}=\delta _{ij}-n_{i}n_{j}\,.  \label{P}
\end{equation}

We find that the computation would be technically quite difficult with the
integral as it is in Eq.~(\ref{h0}): the way the source point $\mathbf{x}%
^{\prime }$ (integration variable) is combined with the field point $\mathbf{%
x}$ in the integrand would make our calculation quite intractable. However,
expressing the integral in expansion, we obtain a more computationally
favorable form:%
\begin{eqnarray}
h_{ij}^{\mathrm{TT}}\left( t,\mathbf{x}\right) &=&\frac{4G}{c^{4}r}\Lambda
_{ij,kl}\left( \mathbf{n}\right) \left[ \int d^{3}x^{\prime }T^{kl}\left( t_{%
\mathrm{R}},\mathbf{x}^{\prime }\right) \right.  \notag \\
&&\left. +\frac{1}{c}n^{m}\frac{d}{dt_{\mathrm{R}}}\int d^{3}x^{\prime
}T^{kl}\left( t_{\mathrm{R}},\mathbf{x}^{\prime }\right) x_{m}^{\prime
}\right.  \notag \\
&&\left. +\frac{1}{2c^{2}}n^{m}n^{p}\frac{d^{2}}{dt_{\mathrm{R}}^{2}}\int
d^{3}x^{\prime }T^{kl}\left( t_{\mathrm{R}},\mathbf{x}^{\prime }\right)
x_{m}^{\prime }x_{p}^{\prime }+\cdots \right] _{t_{\mathrm{R}}=t-r/c}\,,
\label{h1}
\end{eqnarray}%
where $r=\left\vert \mathbf{x}\right\vert $ and $t_{\mathrm{R}}=t-r/c$
denotes the retarded time. In particular, the computation resulting from the
first term alone in the square bracket in Eq. (\ref{h1}) is called the
\textquotedblleft quadrupole approximation\textquotedblright . The next
terms will provide corrections to this computation.

\subsection{Computation of gravitational waves in the quadrupole
approximation\label{quad}}

The complete information about the motion of the colliding two-bubble system
is encoded in the scalar field solution $S=\sqrt{4\pi }\phi $, as given by
Figure~\ref{fig2}, and thus is carried by the energy-momentum tensors
through Eq.~(\ref{T}): to be precise, the energy-momentum tensors are
comprised of the scalar field $S$ and the geometry $g_{\mu \nu }$, which are
obtained by solving Eqs.~(\ref{ein}) and (\ref{scl}) simultaneously~\cite%
{Hwang}. As described by Eqs.~(\ref{h0}) and~(\ref{h1}), GWs from the system
are computed with the energy-momentum tensors being the sources. It is
believed that the two bubbles will be in highly relativistic motion when
they collide~\cite{Hawking}. In view of this, corrections due to the
next-to-leading order terms in Eq.~(\ref{h1}) should not be disregarded if
one aims to compute GWs from the system accurately. However, although not
perfectly accurate, the leading order term alone in Eq.~(\ref{h1}) provides
the \textquotedblleft quadrupole approximation\textquotedblright {}\ of GWs: 
\begin{equation}
_{\mathrm{Q}}h_{ij}^{\mathrm{TT}}\left( t,\mathbf{x}\right) =\frac{4}{r}%
\Lambda _{ij,kl}\left( \mathbf{n}\right) I_{kl}\left( t_{\mathrm{R}}\right)
\,,  \label{Qh}
\end{equation}%
where we have adopted the unit convention $c=G=1$, and%
\begin{equation}
I_{kl}\left( t_{\mathrm{R}}\right) \equiv \int d^{3}x^{\prime }T^{kl}\left(
t_{\mathrm{R}},\mathbf{x}^{\prime }\right) \,.  \label{QI}
\end{equation}%
Throughout this paper our computation is carried out only from this piece.
Our main purpose is to provide \textit{qualitative} insights into patterns
of GWs from the colliding two-bubble system in time domain, and the
next-to-leading order corrections in Eq.~(\ref{h1}) are disregarded in our
analysis.

Following Ref.~\cite{Kosowsky}, we can reduce the amount of computation in a
great deal. As described in Subsection~\ref{basics}, we choose the $x$-axis
to coincide with the line adjoining the centers of the two bubbles. With the
axial symmetry about the $x$-axis, the off-diagonal components are zero and
we can put $I_{kl}$ in the form, 
\begin{equation}
I_{kl}=D\delta_{kl}+\triangle\delta_{kx}\delta_{lx}\,.  \label{T1}
\end{equation}
Here, the first term turns out to be 
\begin{equation}
D=\frac{1}{2}\left(I_{yy}+I_{zz}\right)\,,  \label{D}
\end{equation}
which does not contribute to gravitational radiation due to Eqs.~(\ref{L})
and (\ref{P}). The second term is given by 
\begin{equation}
\triangle=I_{xx}-\frac{1}{2}\left(I_{yy}+I_{zz}\right)\,.  \label{T2}
\end{equation}
Therefore, $I_{kl}$ is practically equivalent to $\triangle\delta_{kx}%
\delta_{lx}$: 
\begin{equation}
I_{kl}\sim\delta_{kx}\delta_{lx}\left[I_{xx}-\frac{1}{2}\left(I_{yy}+I_{zz}%
\right)\right]\,.  \label{al2}
\end{equation}
Then by Eqs.~(\ref{QI}) and (\ref{al2}) we may express 
\begin{equation}
I_{kl}\left(t_{\mathrm{R}}\right)=\delta_{kx}\delta_{lx}\int d^{3}x^{\prime}%
\left[T^{xx}\left(t_{\mathrm{R}},\mathbf{x}^{\prime}\right)-\frac{1}{2} %
\left[T^{yy}\left(t_{\mathrm{R}},\mathbf{x}^{\prime}\right)+T^{zz} \left(t_{%
\mathrm{R}},\mathbf{x}^{\prime}\right)\right]\right]\,.  \label{al3}
\end{equation}

Now, recall from Subsection~\ref{basics} that in the flat spacetime we
define the hyperbolic coordinates $\tau $, $\chi $, $\theta $ by 
\begin{eqnarray}
t_{\mathrm{R}} &=&\tau \cosh \chi \,,  \label{hc1} \\
y &=&\tau \sinh \chi \sin \theta \,,  \label{hc2} \\
z &=&\tau \sinh \chi \cos \theta \,,  \label{hc3}
\end{eqnarray}%
so that 
\begin{equation}
\tau ^{2}=t_{\mathrm{R}}^{2}-\rho ^{2},  \label{hc4}
\end{equation}%
where $0\leq \chi <\infty $, $0\leq \theta <2\pi $ and $\rho \equiv \sqrt{%
y^{2}+z^{2}}$. In these coordinates the flat spacetime metric takes the
form, 
\begin{equation}
ds^{2}=-d\tau ^{2}+dx^{2}+\tau ^{2}\left( d\chi ^{2}+\sinh ^{2}\chi d\theta
^{2}\right) \,.  \label{dsfl}
\end{equation}%
In this geometry, however, the scalar field solution is independent of the
coordinates $\chi $ and $\theta $ and is expressed in the coordinates $%
\left( \tau ,x\right) $ only; namely, $\phi \left( \tau ,x\right) $~\cite%
{Hawking, Kosowsky}. Taking this into account, we should rewrite the volume
element for the integral~(\ref{al3}) as 
\begin{equation}
d^{3}x=dx\rho d\rho d\theta =-dx\tau d\tau d\theta \,,  \label{I4}
\end{equation}%
which is defined at the instant $t_{\mathrm{R}}$ by means of Eq.~(\ref{hc4}).

From Eq.~(\ref{hc4}) we see that $\tau $ has an upper bound $t_{\mathrm{R}}$
with $\rho =0$. This represents the exterior surface of the bubble walls,
i.e.%
\begin{equation}
\tau =t_{\mathrm{R}}.  \label{tau0}
\end{equation}%
However, the interior surface is found from Eq.~(\ref{hc4}) to be 
\begin{equation}
\tau =t_{\mathrm{R}}-\frac{\eta ^{2}}{8t_{\mathrm{R}}}+\mathcal{O}\left( 
\frac{\eta ^{4}}{t_{\mathrm{R}}^{3}}\right) \,,  \label{tau1}
\end{equation}%
given a wall thickness $\eta \sim 2\rho \ll t_{\mathrm{R}}$. Then from Eqs.~(%
\ref{al3}), (\ref{I4}), (\ref{tau0}) and (\ref{tau1}) we can compute $%
I_{kl}\left( t_{\mathrm{R}}\right) $ effectively out of a volume piece $%
\mathcal{V}$: 
\begin{eqnarray}
I_{kl}\left( t_{\mathrm{R}}\right) &=&\delta _{kx}\delta _{lx}\int_{\mathcal{%
V}}d^{3}x^{\prime }\left[ T_{xx}-\frac{1}{2}\left( T_{yy}+T_{zz}\right) %
\right]  \notag \\
&=&2\pi \delta _{kx}\delta _{lx}\int_{t_{\mathrm{R}}-\eta ^{2}/\left( 8t_{%
\mathrm{R}}\right) }^{t_{\mathrm{R}}}d\tau ^{\prime }\tau ^{\prime
}\int_{-x_{\mathrm{o}}}^{x_{\mathrm{o}}}dx^{\prime }\left[ T_{xx}-\frac{1}{2}%
\left( T_{yy}+T_{zz}\right) \right] +\mathcal{O}\left( \frac{\eta ^{4}}{t_{%
\mathrm{R}}^{3}}\right)  \notag \\
&=&\frac{\pi }{4}\delta _{kx}\delta _{lx}\eta ^{2}\int_{-x_{\mathrm{o}}}^{x_{%
\mathrm{o}}}dx^{\prime }\left[ T_{xx}-\frac{1}{2}\left( T_{yy}+T_{zz}\right) %
\right] _{\tau =t_{\mathrm{R}}}+\mathcal{O}\left( \frac{\eta ^{4}}{t_{%
\mathrm{R}}^{3}}\right) \,,  \label{al4}
\end{eqnarray}%
where the volume piece $\mathcal{V}$ is defined from a thin cylindrical
shell in motion, having the thickness $\eta ^{2}/\left( 8t_{\mathrm{R}%
}\right) $, extending along the $x$-axis, by means of Eqs. (\ref{tau0}) and (%
\ref{tau1}): from $t_{\mathrm{R}}-\eta ^{2}/\left( 8t_{\mathrm{R}}\right) +%
\mathcal{O}\left( \eta ^{4}/t_{\mathrm{R}}^{3}\right) \leq \tau \leq $ $t_{%
\mathrm{R}}$ we find $\mathcal{V}=\Delta x\left\vert \tau \Delta \tau
\right\vert \Delta \theta =\frac{\pi }{4}\eta ^{2}\Delta x\left[ 1+\mathcal{O%
}\left( \eta ^{2}/t_{\mathrm{R}}^{2}\right) \right] $, and the limit of the
integral $x_{\mathrm{o}}=\Delta x/2$ should be chosen to be sufficiently
large such that collision effects be fully covered in numerical integration 
\footnote{%
By Eqs.~(\ref{hc4}) and (\ref{I4}) the volume piece can also be viewed as $%
\mathcal{V}=\Delta x\rho \Delta \rho \Delta \theta =\frac{\pi }{4}\eta
^{2}\Delta x\left[ 1+\mathcal{O}\left( \eta ^{2}/t_{\mathrm{R}}^{2}\right) %
\right] $. Then it may be stated that the volume integral in Eq.~(\ref{al4})
will be equivalently evaluated out of this volume piece, whose shape is a
long thin cylinder with the diameter (thickness) $\eta $ surrounding the $x$%
-axis. This is in agreement with the statement from Ref.~\cite{Hawking}:
\textquotedblleft The kinetic energy of the bubble walls will be
concentrated in a small region around the $x$-axis of a wall thickness $\eta 
$.\textquotedblright\ }. Following Ref.~\cite{Hawking}, we estimate a bubble
wall thickness $\eta $, assuming that the walls will be highly relativistic
when they collide, having the Lorentz factor $\gamma $: 
\begin{equation}
\eta \sim \frac{3}{2}\phi _{\mathrm{f}}/\left( \xi ^{2}\gamma \right) \sim 
\frac{3}{2}\phi _{\mathrm{f}}^{2}/\left( b\epsilon ^{4}\right) \,,
\label{thick1}
\end{equation}%
where $\phi _{\mathrm{f}}$ denotes the scalar field value at the false
vacuum and $\xi ^{4}$ the effective height of the potential barrier between
the two minima, and the Lorentz factor $\gamma =b\epsilon ^{4}/\left( \xi
^{2}\phi _{\mathrm{f}}\right) $ with $2b$ representing the separation of the
bubbles and $\epsilon ^{4}$ the potential difference between the two minima
(which is equivalent to $V_{\mathrm{f}}$ in our analysis in Subsection~\ref%
{scalar}) \cite{Hawking}.

However, as described in Subsection~\ref{scalar}, our scalar field $S=\sqrt{%
4\pi }\phi $ is obtained by solving Eqs.~(\ref{ein}) and (\ref{scl1})
simultaneously, using the ansatz given by Eq.~(\ref{dsgen}), in the
coordinates $\left( u,v,\chi ,\theta \right) $. Then by Eq.~(\ref{T}) the
energy-momentum tensors should be expressed in the same coordinates. Now,
due to the definitions of $u$ and $v$ in the flat spacetime limit, and by
Eqs.~(\ref{hc1}), (\ref{hc2}) and (\ref{hc3}) we have%
\begin{eqnarray}
u &=&\tau -x=\sqrt{t_{\mathrm{R}}^{2}-\left( y^{2}+z^{2}\right) }-x\,,
\label{u} \\
v &=&\tau +x=\sqrt{t_{\mathrm{R}}^{2}-\left( y^{2}+z^{2}\right) }+x\,,
\label{v} \\
\chi &=&\tanh ^{-1}\left( \sqrt{\frac{y^{2}+z^{2}}{t_{\mathrm{R}}^{2}}}%
\right) \,.  \label{chi}
\end{eqnarray}%
Using these relations, we find 
\begin{eqnarray}
T_{xx} &=&T_{uu}\left( \frac{\partial u}{\partial x}\right) ^{2}+2T_{uv}%
\frac{\partial u}{\partial x}\frac{\partial v}{\partial x}+T_{vv}\left( 
\frac{\partial v}{\partial x}\right) ^{2}+T_{\chi \chi }\left( \frac{%
\partial \chi }{\partial x}\right) ^{2}  \notag \\
&=&T_{uu}-2T_{uv}+T_{vv},  \label{Ixx} \\
T_{yy} &=&T_{uu}\left( \frac{\partial u}{\partial y}\right) ^{2}+2T_{uv}%
\frac{\partial u}{\partial y}\frac{\partial v}{\partial y}+T_{vv}\left( 
\frac{\partial v}{\partial y}\right) ^{2}+T_{\chi \chi }\left( \frac{%
\partial \chi }{\partial y}\right) ^{2}  \notag \\
&=&\frac{y^{2}}{t_{\mathrm{R}}^{2}-\left( y^{2}+z^{2}\right) }\left(
T_{uu}+2T_{uv}+T_{vv}\right) +\frac{t_{\mathrm{R}}^{2}y^{2}}{\left[ t_{%
\mathrm{R}}^{2}-\left( y^{2}+z^{2}\right) \right] ^{2}\left(
y^{2}+z^{2}\right) }T_{\chi \chi }\,,  \label{Iyy} \\
T_{zz} &=&T_{uu}\left( \frac{\partial u}{\partial z}\right) ^{2}+2T_{uv}%
\frac{\partial u}{\partial z}\frac{\partial v}{\partial z}+T_{vv}\left( 
\frac{\partial v}{\partial z}\right) ^{2}+T_{\chi \chi }\left( \frac{%
\partial \chi }{\partial z}\right) ^{2}  \notag \\
&=&\frac{z^{2}}{t_{\mathrm{R}}^{2}-\left( y^{2}+z^{2}\right) }\left(
T_{uu}+2T_{uv}+T_{vv}\right) +\frac{t_{\mathrm{R}}^{2}z^{2}}{\left[ t_{%
\mathrm{R}}^{2}-\left( y^{2}+z^{2}\right) \right] ^{2}\left(
y^{2}+z^{2}\right) }T_{\chi \chi }\,.  \label{Izz}
\end{eqnarray}

Substituting Eqs.~(\ref{Ixx}), (\ref{Iyy}) and (\ref{Izz}) into Eq.~(\ref%
{al4}), we obtain 
\begin{equation}
I_{kl}\left( t_{\mathrm{R}}\right) =\frac{\pi }{4}\delta _{kx}\delta
_{lx}\eta ^{2}\int_{-x_{\mathrm{o}}}^{x_{\mathrm{o}}}dx^{\prime }\left[
T_{uu}-2T_{uv}+T_{vv}-\frac{1}{2t_{\mathrm{R}}^{2}}T_{\chi \chi }\right]
_{t_{\mathrm{R}}}+\mathcal{O}\left( \frac{\eta ^{4}}{t_{\mathrm{R}}^{3}}%
\right) \,,  \label{Ial}
\end{equation}%
where the subscript $t_{\mathrm{R}}$ outside the square bracket means that
the double-null coordinates $\left( u,v\right) $ are defined at $\tau =t_{%
\mathrm{R}}$; namely, $u=t_{\mathrm{R}}-x$ and $v=t_{\mathrm{R}}+x$. In the
actual computation of Eq.~(\ref{Ial}), we integrate $T_{uu}\left( u,v\right) 
$, $T_{uv}\left( u,v\right) $, $T_{vv}\left( u,v\right) $ and $T_{\chi \chi
}\left( u,v\right) $, which are constructed out of the scalar field solution 
$S\left( u,v\right) =\sqrt{4\pi }\phi \left( u,v\right) $, the geometry
solution $g_{uv}\left( u,v\right) $, $g_{\chi \chi }\left( u,v\right) $, $%
g_{\theta \theta }\left( u,v\right) $ and the potential $V\left( S\right) $
via Eq.~(\ref{T}). Then we need to change the variable of integration, from $%
x$ to $u$ or $v$. Using the relations $u=t_{\mathrm{R}}-x$ and $v=t_{\mathrm{%
R}}+x$, we can convert 
\begin{equation}
dx=-du\text{ \ \ or \ }dx=dv.  \label{var3}
\end{equation}%
Then we may rewrite 
\begin{eqnarray}
\int_{-x_{\mathrm{o}}}^{x_{\mathrm{o}}}dx^{\prime }~T_{ab}\left( u,v\right)
&=&\int_{t_{\mathrm{R}}-x_{\mathrm{o}}}^{t_{\mathrm{R}}+x_{\mathrm{o}%
}}du~T_{ab}\left( u,2t_{\mathrm{R}}-u\right) =\int_{t_{\mathrm{R}}-x_{%
\mathrm{o}}}^{t_{\mathrm{R}}+x_{\mathrm{o}}}dv~T_{ab}\left( 2t_{\mathrm{R}%
}-v,v\right)  \notag \\
&=&2\int_{0}^{x_{\mathrm{o}}}du~T_{ab}\left( t_{\mathrm{R}}+u,t_{\mathrm{R}%
}-u\right) =2\int_{0}^{x_{\mathrm{o}}}dv~T_{ab}\left( t_{\mathrm{R}}-v,t_{%
\mathrm{R}}+v\right) \,,  \label{Ial4}
\end{eqnarray}%
where $T_{ab}$ represents any of $T_{uu}$, $T_{uv}$, $T_{vv}$ and $T_{\chi
\chi }$, and the expressions in the second line have been obtained via
translations, $u\rightarrow u-t_{\mathrm{R}}$ and $v\rightarrow v-t_{\mathrm{%
R}}$. Then by Eqs.~(\ref{Ial}) and (\ref{Ial4}) $I_{kl}\left( t_{\mathrm{R}%
}\right) $ can be expressed as 
\begin{eqnarray}
I_{kl}\left( t_{\mathrm{R}}\right) &=&\frac{\pi }{2}\delta _{kx}\delta
_{lx}\eta ^{2}\left[ \int_{0}^{x_{\mathrm{o}}}du~T_{uu}\left( t_{\mathrm{R}%
}\pm u,t_{\mathrm{R}}\mp u\right) -2\int_{0}^{x_{\mathrm{o}}}du~T_{uv}\left(
t_{\mathrm{R}}\pm u,t_{\mathrm{R}}\mp u\right) \right.  \notag \\
&&\!\!\left. +\int_{0}^{x_{\mathrm{o}}}du~T_{vv}\left( t_{\mathrm{R}}\pm
u,t_{\mathrm{R}}\mp u\right) -\frac{1}{2t_{\mathrm{R}}^{2}}\int_{0}^{x_{%
\mathrm{o}}}du~T_{\chi \chi }\left( t_{\mathrm{R}}\pm u,t_{\mathrm{R}}\mp
u\right) \right] +\mathcal{O}\left( \frac{\eta ^{4}}{t_{\mathrm{R}}^{3}}%
\right) \,.  \label{Iab}
\end{eqnarray}

If the wall thickness $\eta $ can be taken sufficiently small in Eq.~(\ref%
{Iab}), then by Eq.~(\ref{Qh}) we can compute the bubble-collision-induced
GWs in the quadrupole approximation as 
\begin{eqnarray}
_{\mathrm{Q}}h_{ij}^{\mathrm{TT}}\left( t,\mathbf{x}\right) &\approx &\frac{%
2\pi }{r}\eta ^{2}\Lambda _{ij,kl}\left( \mathbf{n}\right) \delta
_{kx}\delta _{lx}  \notag \\
&&\times \text{ }\left[ \int_{0}^{x_{\mathrm{o}}}du~T_{uu}\left( t_{\mathrm{R%
}}\pm u,t_{\mathrm{R}}\mp u\right) -2\int_{0}^{x_{\mathrm{o}%
}}du~T_{uv}\left( t_{\mathrm{R}}\pm u,t_{\mathrm{R}}\mp u\right) \right. 
\notag \\
&&\left. +\int_{0}^{x_{\mathrm{o}}}du~T_{vv}\left( t_{\mathrm{R}}\pm u,t_{%
\mathrm{R}}\mp u\right) -\frac{1}{2t_{\mathrm{R}}^{2}}\int_{0}^{x_{\mathrm{o}%
}}du~T_{\chi \chi }\left( t_{\mathrm{R}}\pm u,t_{\mathrm{R}}\mp u\right) %
\right] \,.  \label{Ial5}
\end{eqnarray}%
Now, without loss of generality we may choose 
\begin{equation}
\mathbf{n=}{\left( n_{x},n_{y},n_{z}\right) =\left( \cos \vartheta ,\sin
\vartheta ,0\right) }\,,  \label{n}
\end{equation}%
where $\vartheta $ denotes the angle of propagation taken from the $x$-axis.
From this it follows that 
\begin{equation}
\Lambda _{ij,kl}\left( \mathbf{n}\right) \delta _{kx}\delta _{lx}=\delta
_{ix}\delta _{jx}-2\delta _{ix}n_{j}\cos \vartheta +\frac{1}{2}%
n_{i}n_{j}\left( 1+\cos ^{2}\vartheta \right) -\frac{1}{2}\delta _{ij}\sin
^{2}\vartheta \,,  \label{L1}
\end{equation}%
due to Eqs.~(\ref{L}) and (\ref{P}). Substituting this into Eq.~(\ref{Ial5}%
), we finally express 
\begin{eqnarray}
_{\mathrm{Q}}h_{ij}^{\mathrm{TT}}\left( t,\mathbf{x}\right) &\approx &\frac{%
2\pi }{r}\eta ^{2}\left[ \delta _{ix}\delta _{jx}-2\delta _{ix}n_{j}\cos
\vartheta +\frac{1}{2}n_{i}n_{j}\left( 1+\cos ^{2}\vartheta \right) -\frac{1%
}{2}\delta _{ij}\sin ^{2}\vartheta \right]  \notag \\
&&\times \text{ }\left[ \int_{0}^{x_{\mathrm{o}}}du~T_{uu}\left( t_{\mathrm{R%
}}\pm u,t_{\mathrm{R}}\mp u\right) -2\int_{0}^{x_{\mathrm{o}%
}}du~T_{uv}\left( t_{\mathrm{R}}\pm u,t_{\mathrm{R}}\mp u\right) \right. 
\notag \\
&&\left. +\int_{0}^{x_{\mathrm{o}}}du~T_{vv}\left( t_{\mathrm{R}}\pm u,t_{%
\mathrm{R}}\mp u\right) -\frac{1}{2t_{\mathrm{R}}^{2}}\int_{0}^{x_{\mathrm{o}%
}}du~T_{\chi \chi }\left( t_{\mathrm{R}}\pm u,t_{\mathrm{R}}\mp u\right) %
\right] \,,  \label{Ial6}
\end{eqnarray}%
where $t_{\mathrm{R}}=t-r=t-\left\vert \mathbf{x}\right\vert $, and the wall
thickness $\eta $ can be specified by means of Eq. (\ref{thick1}); namely,
in terms of the quantities for the bubble collision profiles, such as the
false vacuum field $\phi _{\mathrm{f}}$ (equivalent to $S_{\mathrm{f}}/\sqrt{%
4\pi }$), the potential difference between the two minima $\epsilon ^{4}$
(equivalent to $V_{\mathrm{f}}$) and half the separation of the bubbles $b$~%
\cite{Hawking}.

\begin{figure}[t]
\begin{center}
\includegraphics[scale=0.22]{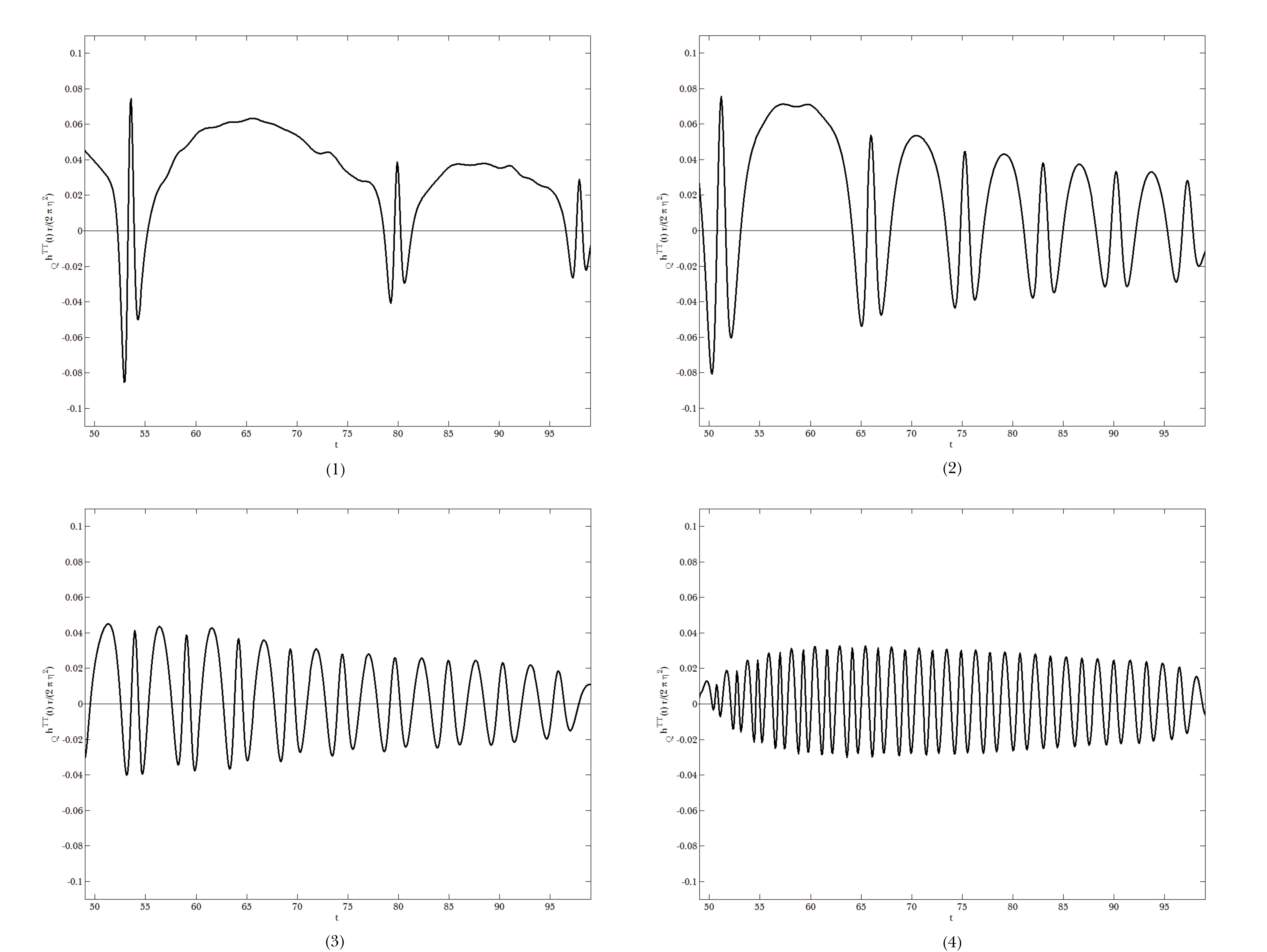}
\end{center}
\caption{The numerical plots of $_{\mathrm{Q}}h^{\mathrm{TT}%
}\left(t\right)r/\left(2\protect\pi\protect\eta^{2}\right)$\ with various
false vacuum field values, $S_{\mathrm{f}}=\protect\sqrt{4\protect\pi}%
\protect\phi_{\mathrm{f}}=$ (1) $0.1$, (2) $0.2$, (3) $0.3$, (4) $0.4$. As $%
\protect\eta^{2}\sim\protect\phi_{\mathrm{f}}^{4}$, the amplitude of $_{%
\mathrm{Q}}h^{\mathrm{TT}}\left(t\right)$ should scale as (1) $1$, (2) $%
2^{4} $, (3) $3^{4}$, (4) $4^{4}$.}
\label{fig4}
\end{figure}

\textbf{RESULT 1:} The numerical computations of Eq. (\ref{Ial6}) are
presented in Figure~\ref{fig4}; with various false vacuum field values, $S_{%
\mathrm{f}}=\sqrt{4\pi }\phi _{\mathrm{f}}=$ (1) $0.1$, (2) $0.2$, (3) $0.3$%
, (4) $0.4$, in accordance with the scalar field solutions as presented by
Figure~\ref{fig2}. Due to Eqs.~(\ref{thick1}) and (\ref{Ial6}), the
amplitude of our GWs $_{\mathrm{Q}}h^{\mathrm{TT}}\left( t\right) $ scales
as $\phi _{\mathrm{f}}^{4}$ if the other conditions, $\epsilon ^{4}$ and $b$
are kept the same. Thus, with $S_{\mathrm{f}}=$ (1) $0.1$, (2) $0.2$, (3) $%
0.3$, (4) $0.4$, the amplitude scales as (1) $1$, (2) $2^{4}$, (3) $3^{4}$,
(4) $4^{4}$. The frequency of the waves is modulating due to the
non-linearity of the collision dynamics in the all four cases of $S_{\mathrm{%
f}}$, (1) - (4). However, the modulating frequency increases overall as $S_{%
\mathrm{f}}$ increases, which is analogous to the{\ tendency exhibited by }$%
S\left( u,v\right) =\sqrt{4\pi }\phi \left( u,v\right) ${\ as shown in
Figure~\ref{fig2}. }In Figure~\ref{fig4}, we present $_{\mathrm{Q}}h^{%
\mathrm{TT}}\left( t\right) r/\left( 2\pi \eta ^{2}\right) $ instead of $_{%
\mathrm{Q}}h^{\mathrm{TT}}\left( t\right) $, and thus all the waveforms are
plotted in the same scale. One should note here that our actual numerical
data of the energy-momentum tensors $T_{ab}$ for Eq.~(\ref{Ial6}) have been
obtained via Eq.~(\ref{T}) after solving Eqs.~(\ref{ein}) and (\ref{scl})
simultaneously~\cite{Hwang}. Therefore, our $T_{ab}$ contain the full
physical information about the bubble collisions in terms of the scalar
field $S=\sqrt{4\pi }\phi $, the geometry $g_{ab}$ and the potential $%
V\left( S\right) $; with the radiation reaction effects included in $S$ and $%
g_{ab}$ \footnote{%
The way our GWs are calculated here resembles a \textquotedblleft
semi-relativistic treatment\textquotedblright ,\ originated by Ruffini and
Sasaki~\cite{Ruffini}, in the following senses: (a) the field ($h_{ab}$)
radiates as if it were in flat spacetime, (b) the source ($T_{ab}$) contains
the General Relativistic information about its local spacetime. In Eq. (\ref%
{h0}) we see that our GWs $h_{ab}$ result from distant sources $T_{ab}$,
which are composed of the scalar field and the local geometry given via Eqs.
(\ref{T}), (\ref{dsgen}) and (\ref{scl1}), thus containing the full General
Relativistic information, including the radiation reaction effects.}.

\subsection{A simplified method to compute gravitational waves in the
quadrupole approximation\label{alter}}

Ref.~\cite{Kosowsky} presents a simplified method to compute the GWs $_{%
\mathrm{Q}}h_{ij}^{\mathrm{TT}}\left( t,\mathbf{x}\right) $ of Eq.~(\ref{Qh}%
) by neglecting the gravitational effects on the bubbles: namely, $g_{\mu
\nu }$ in Eq.~(\ref{T}) is replaced by $\eta _{\mu \nu }$, assuming that the
bubbles are in flat spacetime. Then Eq.~(\ref{QI}) can be simplified as 
\begin{equation}
I_{kl}\left( t_{\mathrm{R}}\right) =\int d^{3}x^{\prime }T^{kl}\left( t_{%
\mathrm{R}},\mathbf{x}^{\prime }\right) =\int d^{3}x^{\prime }\partial
_{k}\phi \left( t_{\mathrm{R}},\mathbf{x}^{\prime }\right) \partial _{l}\phi
\left( t_{\mathrm{R}},\mathbf{x}^{\prime }\right) \,,  \label{I1}
\end{equation}%
where the energy-momentum tensors from Eq.~(\ref{T}) have been reduced; $%
T_{ij}\rightarrow \partial _{i}\phi \partial _{j}\phi $ because the terms
proportional to $\delta _{ij}$ in $T_{ij}$ makes no contribution to
gravitational radiation in Eq.~(\ref{Qh}) due to the property of Eq.~(\ref{L}%
); namely, $\Lambda _{ij,kl}\delta _{ij}=0$~\cite{Kosowsky}.

By Eqs.~(\ref{al2}) and (\ref{I1}) we have 
\begin{eqnarray}
I_{kl} &=&\delta _{kx}\delta _{lx}\int d^{3}x^{\prime }\left[ \left( \frac{%
\partial \phi }{\partial x^{\prime }}\right) ^{2}-\frac{1}{2}\left( \frac{%
\partial \phi }{\partial y^{\prime }}\right) ^{2}-\frac{1}{2}\left( \frac{%
\partial \phi }{\partial z^{\prime }}\right) ^{2}\right]  \notag \\
&=&\delta _{kx}\delta _{lx}\int d^{3}x^{\prime }\left[ \left( \frac{\partial
\phi }{\partial x^{\prime }}\right) ^{2}-\frac{1}{2}\left( \frac{\partial
\phi }{\partial \rho ^{\prime }}\right) ^{2}\right] \,.  \label{I2}
\end{eqnarray}%
Using Eq.~(\ref{hc4}), we can modify 
\begin{equation}
\left( \frac{\partial \phi }{\partial \rho }\right) ^{2}=\left( \frac{%
\partial \phi }{\partial \tau }\right) ^{2}\left( \frac{\partial \tau }{%
\partial \rho }\right) ^{2}=\frac{t_{\mathrm{R}}^{2}-\tau ^{2}}{\tau ^{2}}%
\left( \frac{\partial \phi }{\partial \tau }\right) ^{2}\,.  \label{I3}
\end{equation}%
Then in the same manner as described above by Eq.~(\ref{al4}), the integral $%
I_{kl}\left( t_{\mathrm{R}}\right) $ is computed out of the volume piece $%
\mathcal{V}=\Delta x\left\vert \tau \Delta \tau \right\vert \Delta \theta =%
\frac{\pi }{4}\eta ^{2}\Delta x\left[ 1+\mathcal{O}\left( \eta ^{2}/t_{%
\mathrm{R}}^{2}\right) \right] $: 
\begin{eqnarray}
I_{kl}\left( t_{\mathrm{R}}\right) &=&\delta _{kx}\delta _{lx}\int_{\mathcal{%
V}}d^{3}x^{\prime }\left[ \left( \frac{\partial \phi }{\partial x^{\prime }}%
\right) ^{2}-\frac{1}{2}\left( \frac{\partial \phi }{\partial \rho ^{\prime }%
}\right) ^{2}\right]  \notag \\
&=&2\pi \delta _{kx}\delta _{lx}\int_{t_{\mathrm{R}}-\eta ^{2}/\left( 8t_{%
\mathrm{R}}\right) }^{t_{\mathrm{R}}}d\tau ^{\prime }\tau ^{\prime
}\int_{-x_{\mathrm{o}}}^{x_{\mathrm{o}}}dx^{\prime }\left[ \left( \frac{%
\partial \phi }{\partial x^{\prime }}\right) ^{2}-\frac{t_{\mathrm{R}%
}^{2}-\tau ^{\prime 2}}{2\tau ^{\prime 2}}\left( \frac{\partial \phi }{%
\partial \tau ^{\prime }}\right) ^{2}\right] +\mathcal{O}\left( \frac{\eta
^{4}}{t_{\mathrm{R}}^{3}}\right)  \notag \\
&=&\frac{\pi }{2}\delta _{kx}\delta _{lx}\eta ^{2}\int_{0}^{x_{\mathrm{o}%
}}dx^{\prime }\left( \frac{\partial \phi }{\partial x^{\prime }}\right)
_{\tau =t_{\mathrm{R}}}^{2}+\mathcal{O}\left( \frac{\eta ^{4}}{t_{\mathrm{R}%
}^{3}}\right) \,.  \label{I5}
\end{eqnarray}%
If the wall thickness $\eta $ can be taken sufficiently small in Eq.~(\ref%
{I5}), by Eqs.~(\ref{Qh}) and (\ref{I5}) we can compute the
bubble-collision-induced GWs in the quadrupole approximation as 
\begin{equation}
_{\mathrm{Q}}h_{ij}^{\mathrm{TT}}\left( t,\mathbf{x}\right) \approx \frac{%
2\pi }{r}\Lambda _{ij,kl}\left( \mathbf{n}\right) \delta _{kx}\delta
_{lx}\eta ^{2}\int_{0}^{x_{\mathrm{o}}}dx^{\prime }\left( \frac{\partial
\phi }{\partial x^{\prime }}\right) _{\tau =t_{\mathrm{R}}}^{2}\,.
\label{Ih}
\end{equation}%
Substituting Eq.~(\ref{L1}) into Eq.~(\ref{Ih}), we finally express 
\begin{eqnarray}
_{\mathrm{Q}}h_{ij}^{\mathrm{TT}}\left( t,\mathbf{x}\right) &\approx &\frac{%
2\pi }{r}\eta ^{2}\left[ \delta _{ix}\delta _{jx}-2\delta _{ix}n_{j}\cos
\vartheta +\frac{1}{2}n_{i}n_{j}\left( 1+\cos ^{2}\vartheta \right) -\frac{1%
}{2}\delta _{ij}\sin ^{2}\vartheta \right]  \notag \\
&&\times \int_{0}^{x_{\mathrm{o}}}dx^{\prime }\left( \frac{\partial \phi }{%
\partial x^{\prime }}\right) _{\tau =t_{\mathrm{R}}}^{2}\,,  \label{h3}
\end{eqnarray}%
where $t_{\mathrm{R}}=t-r=t-\left\vert \mathbf{x}\right\vert $, and $\eta $
is specified by Eq. (\ref{thick1}).

\begin{figure}[tbp]
\begin{center}
\includegraphics[scale=0.4]{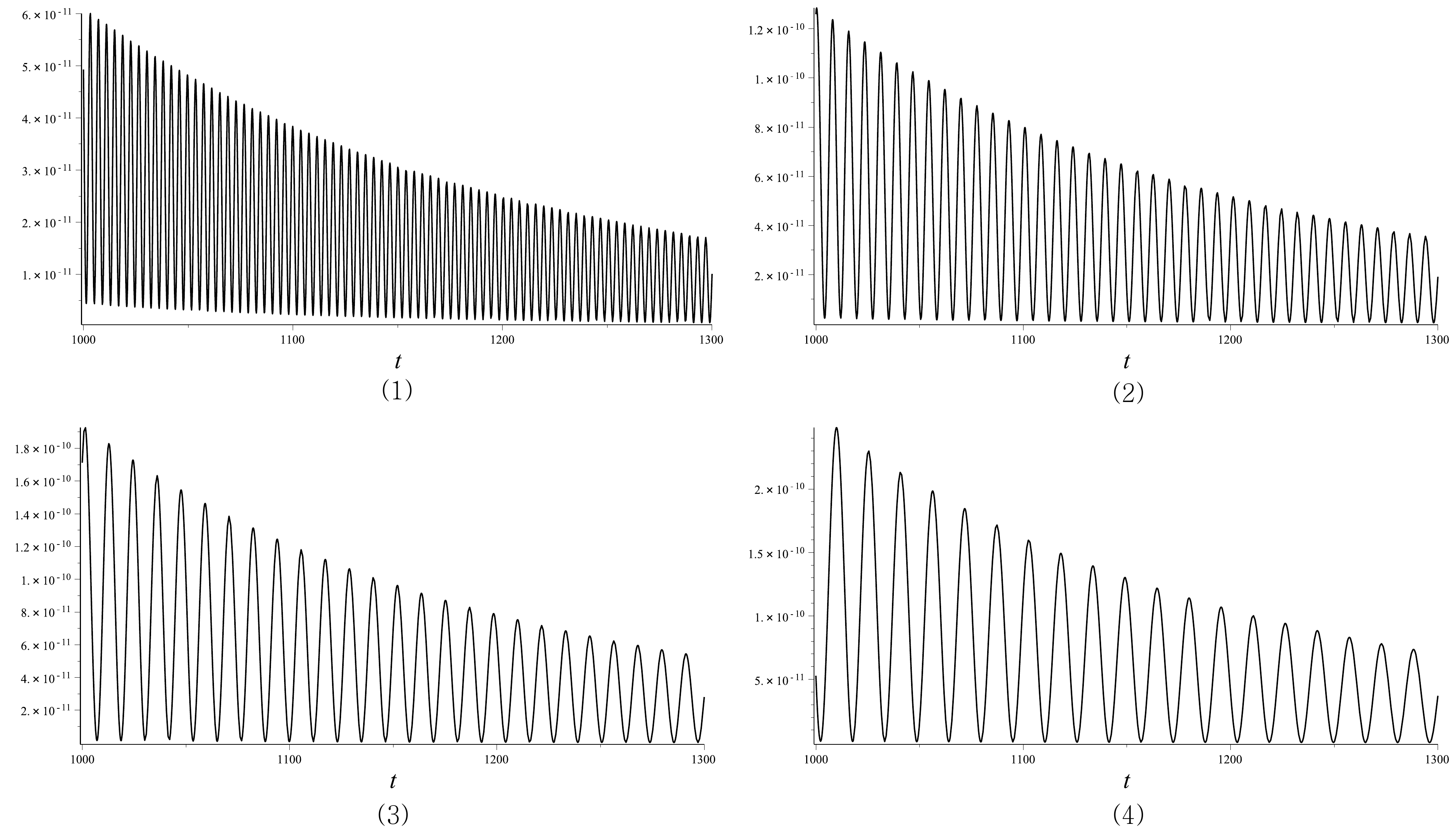}
\end{center}
\caption{Plots of $_{\mathrm{Q}}h^{\mathrm{TT}}\left( t\right) r/\left( 2%
\protect\pi \protect\eta ^{2}\right) $ computed via the expression in Eq.~(%
\protect\ref{h3}), with $\protect\phi $ given by Eq. (\protect\ref{phi-appr}%
). Corresponding to the false vacuum field values, $S_{\mathrm{f}}=\protect%
\sqrt{4\protect\pi }\protect\phi _{\mathrm{f}}=$ (1) $0.1$, (2) $0.2$, (3) $%
0.3$, (4) $0.4$, are the frequencies, $\protect\omega _{\mathrm{t}}\simeq $
(1) $0.8124$, (2) $0.4062$, (3) $0.2708$, (4) $0.2031$ $\sim $ (1) $1$, (2) $%
1/2$, (3) $1/3$, (4) $1/4$. And the amplitude of $_{\mathrm{Q}}h^{\mathrm{TT}%
}\left( t\right) r/\left( 2\protect\pi \protect\eta ^{2}\right) $ scales as
(1) $1$, (2) $2$, (3) $3$, (4) $4$ due to the factor $\protect\omega _{%
\mathrm{t}}^{-1}\sim S_{\mathrm{f}}$. Then the amplitude of $_{\mathrm{Q}}h^{%
\mathrm{TT}}\left( t\right) $ should scale as $\protect\eta ^{2}\protect%
\omega _{\mathrm{t}}^{-1}\sim S_{\mathrm{f}}^{5}\sim $ (1) $1$, (2) $2^{5}$,
(3) $3^{5}$, (4) $4^{5}$.}
\label{fig5}
\end{figure}

\textbf{RESULT 2:} Toward the end of the bubble collisions, $\tau \gg 1$,
the scalar field oscillates around the true vacuum state, i.e. $\left\vert
\phi \right\vert \ll 1$, being nearly \textit{monochromatic}. Then we can
approximate Eq.~(\ref{scl1}) as%
\begin{equation}
\square \phi \left( \tau ,x\right) \approx \frac{6\left( \beta +1\right) V_{%
\mathrm{f}}}{\beta S_{\mathrm{f}}^{2}}\phi \left( \tau ,x\right) \,,
\label{scl-appr}
\end{equation}%
where we have replaced the curved spacetime Laplacian $\nabla ^{2}$ by the
flat spacetime d'Alembertian $\square \equiv -\partial ^{2}/\partial \tau
^{2}-\left( 2/\tau \right) \partial /\partial \tau +\partial ^{2}/\partial
x^{2}$, neglecting the gravitational effects on the bubbles to simplify the
problem \footnote{%
A similar analysis is found in Ref.~\cite{Johnson}, in which the scalar
field equation is solved in hyperbolic \textquotedblleft de
Sitter\textquotedblright {}\ spacetime in the limit, $\tau \gg H^{-1}$,
where $H$ is the Hubble parameter. The solution shows fluctuations of
decreasing amplitude and increasing period (or decreasing frequency) in $%
\tau $. However, in our analysis, the equation is solved in hyperbolic
\textquotedblleft flat\textquotedblright {}\ spacetime and our solution
given by Eq. (\ref{phi-appr}) has fluctuations of decreasing amplitude and
fixed period (or single frequency; monochromatic) in $\tau $.}. With the
help of Ref.~\cite{Polyanin}, we obtain a solution for Eq.~(\ref{scl-appr}): 
\begin{equation}
\phi \left( \tau ,x\right) =\phi _{\mathrm{o}}\frac{J_{1}\left( \omega _{%
\mathrm{t}}\sqrt{\tau ^{2}-x^{2}}\right) }{\omega _{\mathrm{t}}\sqrt{\tau
^{2}-x^{2}}}\,,  \label{phi-appr}
\end{equation}%
where $J_{n}$ denotes the Bessel function of the first kind, $\omega _{%
\mathrm{t}}\equiv \sqrt{6\left( \beta +1\right) V_{\mathrm{f}}/\left( \beta
S_{\mathrm{f}}^{2}\right) }$ represents the `terminal' frequency of the
bubble collisions, and $\phi _{\mathrm{o}}$ is the amplitude which is
determined by the initial conditions of the field. Substituting Eq.~(\ref%
{phi-appr}) into Eq.~(\ref{h3}), the GWs emitted from the bubble collisions
in the final stage can be computed. Figure \ref{fig5} shows the GWs $_{%
\mathrm{Q}}h^{\mathrm{TT}}\left( t\right) r/\left( 2\pi \eta ^{2}\right) $
computed with various false vacuum field values, $S_{\mathrm{f}}=\sqrt{4\pi }%
\phi _{\mathrm{f}}=$ (1) $0.1$, (2) $0.2$, (3) $0.3$, (4) $0.4$.
Corresponding to the field values are the frequencies, $\omega _{\mathrm{t}%
}\simeq $ (1) $0.8124$, (2) $0.4062$, (3) $0.2708$, (4) $0.2031$ $\sim $ (1) 
$1$, (2) $1/2$, (3) $1/3$, (4) $1/4$, as can be seen from Figure \ref{fig5}.
Also, the amplitude of $_{\mathrm{Q}}h^{\mathrm{TT}}\left( t\right) r/\left(
2\pi \eta ^{2}\right) $ scales as (1) $1$, (2) $2$, (3) $3$, (4) $4$ due to
the factor $\omega _{\mathrm{t}}^{-1}\sim S_{\mathrm{f}}$, as can be seen
from Figure \ref{fig5}. Then the amplitude of $_{\mathrm{Q}}h^{\mathrm{TT}%
}\left( t\right) $ should scale as $\eta ^{2}\omega _{\mathrm{t}}^{-1}\sim
S_{\mathrm{f}}^{5}\sim $ (1) $1$, (2) $2^{5}$, (3) $3^{5}$, (4) $4^{5}$.

\section{Conclusions}

We have computed GWs emitted from collisions of two equal-sized bubbles in
time domain. The waveforms have been obtained for (i) the
initial-to-intermediate stage of strong collisions and (ii) the final stage
of weak collisions, in \textit{full General Relativity} and in the flat
spacetime approximation, using numerical and analytical methods,
respectively. During (i), the waveforms show the non-linearity of the
collisions, characterized by a modulating frequency and cusp-like bumps,
whereas during (ii), the waveforms exhibit the linearity of the collisions,
featured by a constant frequency and smooth oscillations, as can be checked
from Figures~\ref{fig4} and \ref{fig5}, respectively. Also, depending on the
false vacuum field value $\phi_{\mathrm{f}}$, the waveforms have different
scales of frequency. During (i), the modulating frequency increases overall
as the false vacuum field value $\phi _{\mathrm{f}}$ increases, whereas
during (ii), the frequency $\omega _{\mathrm{t}}$ decreases as the false
vacuum field value $\phi _{\mathrm{f}}$ increases in an inversely
proportional relationship, i.e. $\omega _{\mathrm{t}}\sim \phi _{\mathrm{f}%
}^{-1}$. It is interesting to note that the relationship between the false
vacuum field value and the frequency during (i) changes almost inversely
during (ii). In addition, the false vacuum field value $\phi _{\mathrm{f}}$
affects the amplitude of the waveforms. During (i), the amplitude scales as $%
\eta ^{2}\sim \phi _{\mathrm{f}}^{4}$, whereas during (ii), the amplitude
scales as $\eta ^{2}\omega _{\mathrm{t}}^{-1}\sim \phi _{\mathrm{f}}^{5}$,
where $\eta $ is a bubble wall thickness.

One of the notable differences between the waveforms emitted during (i) and
during (ii) is the sign, as can be seen from Figures~\ref{fig4} and \ref%
{fig5}. This is due to the difference between Eqs.~(\ref{Ial6}) and (\ref{h3}%
): the integral in Eq. (\ref{h3}) is always positive while its counterpart
in Eq. (\ref{Ial6}) is not necessarily. This\ has to do with the composition
of the integrands in the two expressions. The integrand in Eq. (\ref{Ial6})
consists of the energy-momentum tensors $T_{ab}$ which have been obtained
via Eq.~(\ref{T}) after solving Eqs.~(\ref{ein}) and (\ref{scl})
simultaneously~\cite{Hwang}: thus $T_{ab}$ contain the full physical
information of bubble collisions in terms of the scalar field $\phi $, the
geometry $g_{ab}$ and the potential $V\left( \phi \right) $; with the
radiation reaction effects included in $\phi $ and $g_{ab}$. However, as
explained in the beginning of Subsection~\ref{alter}, the integrand in Eq. (%
\ref{h3}) comes only from the first term, with the second and third terms
being disregarded in Eq. (\ref{T}) as the gravitational effects on the
bubbles are assumed to be neglected, following Ref.~\cite{Kosowsky}. This,
combined with the thin-wall approximation, results in the integrand in Eq. (%
\ref{h3}) being positive, which leads to the integral being also positive.
But this is not the case for the integral in Eq. (\ref{Ial6}) due to the
minus signs appearing in Eq. (\ref{T}) and in the integrand in Eq. (\ref%
{Ial6}).

Throughout the paper, we used the thin-wall and quadrupole approximations to
simplify our computations. These approximations served our purpose well in
that we were able to gain some \textit{qualitative} insights into the
time-domain gravitational waveforms emitted from bubble collisions. However,
to obtain more physically reasonable waveforms, taking into account a
generic thickness and relativistic motion of bubble wall, it will be
inevitable to include in our computations the next-to-leading order
corrections beyond each approximation. Huge amount of computation will be
involved in this task, and we leave it for follow-up studies.

\section*{Acknowledgments}

The authors would like to thank Dong-il Hwang for his valuable comments and
assistance during an early stage of this work. The authors also would like
to thank Hongsu Kim, Sang Pyo Kim, Hyung Won Lee, Gungwon Kang and Inyong
Cho for fruitful discussions and helpful comments. BHL, WL and DY appreciate
Pauchy W. Y. Hwang and Sang Pyo Kim for their hospitality at the 9th
International Symposium on Cosmology and Particle Astrophysics in Taiwan,
13-17 November, 2012. DHK and WL appreciate APCTP for its hospitality during
completion of this work. DHK and JY were supported by Basic Science Research
Program through the National Research Foundation of Korea (NRF) funded by
the Ministry of Education (2013R1A1A2008901 and 2013R1A1A2A10004883). BHL
was supported by the National Research Foundation of Korea (NRF) grant
funded by the Korea government (MSIP) (2014R1A2A1A01002306). WL was
supported by Basic Science Research Program through the National Research
Foundation of Korea (NRF) funded by the Ministry of Education
(2012R1A1A2043908). DY was supported by the JSPS Grant-in-Aid for Scientific
Research (A) (No. 21244033) and also supported by Leung Center for Cosmology
and Particle Astrophysics (LeCosPA) of National Taiwan University (103R4000).

\end{document}